\documentstyle[aps,twocolumn,epsf]{revtex}

\newcommand{\be}{\begin{eqnarray}}
\newcommand{\ee}{\end{eqnarray}}
\newcommand{\no}{\nonumber}

\begin{document}

\title{Three ideas on magnetic mass}

\author{V.P. Nair}

\address{Physics Department, City College of the CUNY, New York, NY 10031, USA}

\maketitle

\begin{abstract}

Three topics, the self-consistent resummation of the perturbative expansion for thermal
Yang-Mills (YM) theory, nonperturbative analysis of Yang-Mills theories in (2+1) dimensions
and modification of the Bose distribution for gluons, are discussed. These topics are
related to the magnetic mass or magnetic screening in the quark-gluon plasma.
The value of the magnetic mass is argued to be close to $e^2N/2\pi=g^2TN/2\pi$ for an $SU(N)$
gauge theory. An analytic calculation of the string tension for $YM_{2+1}$, which agrees with
numerical simulations to within $3\%$, is another result.

\end{abstract}


\narrowtext


\section{Introduction}
\label{sec1}

\def \bA {{\bar A}}
\def \bdel {{\bar \partial}}
\def \d {{\delta}}
\def \del {{\partial}}
\def \E {{\cal E}}
\def \H {{\cal H}}
\def \half {{\textstyle{1\over 2}}}
\def \la {{\langle}}
\def \O {{\cal O}}
\def \ra {{\rangle}}
\def \Tr {{\rm Tr}}
\def \vf {{\varphi}}
\def \vk {{\vec k}}
\def \vu {{\vec u}}
\def \vv {{\vec v}}
\def \vx {{\vec x}}
\def \vy {{\vec y}}
\def \S {{\cal S}}

In this talk, I shall discuss some ideas on how a magnetic mass can
be dynamically  generated in a nonabelian gauge theory.
The ultimate
goal, of course, is to understand the thermal perturbative expansion of 
Yang-Mills (YM) theory, at least to the same extent as the zero-temperature
theory. 
Let me start with some simple, well known, observations about the magnetic mass. 
\vskip .1in
\noindent{\it Why do we need a magnetic mass ? }
\vskip .1in
It is well known
that the thermal perturbative expansion of Yang-Mills theory suffers from infrared
divergences.
One set of such divergences are associated with the so-called hard thermal loops
\cite{hardtherm};
these are of the ``electric type" in the sense that they
can be cured by taking account of the
electric screening mass. The hard thermal loop effective action, in fact,
gives a gauge-invariant
definition of the electric mass and eventually a systematic way of reorganizing
perturbation theory to include the electric screening. However, the
perturbative expansion also contains other, ``magnetic type", of divergences which can be
eliminated only if there is screening of magnetostatic interactions, 
or in other words, if there is a magnetic mass \cite{lingpy}.
\vskip .1in
\noindent{\it How can magnetic mass arise in Yang-Mills theories ? }
\vskip .1in
At a very
heuristic level, it is easy to see why magnetic screening is to be expected.
Nonabelian gauge theories have magnetic monopole-like field configurations
which are thermally excited and such a monopole-antimonopole plasma can screen
magnetic interactions just as the charges of the electrical type can screen 
electrostatic interactions.
A more detailed argument can be made as follows. In the imaginary time
formalism, with Matsubara frequencies $\omega_n = 2\pi n T$, where $T$ is the
temperature, the gauge fields have a mode expansion as
$A_i(\vx ,x^0) = \sum_n A_{i,n}(\vx ) \exp ({i2\pi n T x^0})$. At high temperatures and for
modes of wavelength long compared to $1/T$, the modes with nonzero Matsubara frequencies are
unimportant and the theory reduces to the theory of the $\omega_n =0$ mode, viz., a
three (Euclidean) dimensional Yang-Mills theory (or a (2+1)-dimensional theory in
a Wick rotated version). Yang-Mills theories in three or (2+1) dimensions are expected to have
a mass gap and this is effectively the magnetic mass of the (3+1)-dimensional
theory at high temperature \cite{lingpy}. 
This gives the qualitative origin and a quantitative
first approximation to the magnetic mass. The $\omega_n \neq 0$
modes can give small corrections to this mass. 

Many indirect
arguments have been proposed for the existence of a mass gap in $YM_3$ or
$YM_{2+1}$. For example, we can consider an $SU(2)$ gauge theory spontaneously broken
to $U(1)$. This is effectively compact electrodynamics which has a mass gap.
This feature could survive in the unbroken theory if one has sufficient
smoothness as the parameters are relaxed towards the unbroken phase.
The recent conjectured connexion between supergravity on anti-de Sitter (AdS) space
and gauge theories on the boundary of this AdS space gives another way to,
at least qualitatively, argue that there should be a mass gap \cite{mald}.
\vskip .1in
\noindent{\it Why is magnetic mass important ?}
\vskip .1in
The individual Feynman diagrams of thermal perturbation theory are divergent
and therefore, for meaningful calculations, we need the screening masses.
The best definition of continuum Yang-Mills theory at zero temperature 
is in terms of its perturbative expansion. There is a systematic way to
obtain finite meaningful results for this case. At least, term by term, the expansion
is well defined;
questions of convergence, Borel summability, etc. are another matter. 
Ultimately, apart from questions of calculability, there is the question of
principle: Does thermal
perturbative YM theory exist at least to the same extent as the zero temperature limit exists ?
This is important since it may be the only way to define thermal YM theory.
(A lattice definition is more difficult for the thermal case, especially
for nonequilibrium situations. If we have a well-defined
continuum theory, it can be extended, with some labor, to the
nonequilibrium situations.)

I shall discuss three approaches: the self-consistent resummation of perturbation
theory, a nonperturbative analysis of the mass gap in $YM_{2+1}$ (which is
also very interesting in its own right) and the possibility of modifying the 
thermal distribution
to take account of screening. There are a number of other approaches, 
including lattice
based ones, which
have been tried and which I shall not discuss here \cite{general}. 
(See O. Philipsen's article for
a discussion partly complementary to mine.)

\section{Self-consistent resummation} 
\label{sec2}


In this approach, we reorganize the perturbation expansion by adding and subtracting 
a mass term
to the action \cite{NAN,BP,JP1}. 
The general strategy is as follows. We write the action as
\be
\S = \S_{YM} +m^2 \S_m ~-~ \Delta \S_m
\ee
$\S_{YM}$ is the standard Yang-Mills part of the action and $\S_m$ is a mass term
which
must respect gauge invariance. $\Delta$ is a parameter which is taken to have
a loop expansion, viz., $\Delta =\Delta^{(1)} +\Delta^{(2)}+\cdots$,
$\Delta^{(i)}$ being the $i$-th loop term.
Calculations will be done in a loop expansion wherein $\S_{YM}+m^2 \S_m$ is taken to
be the zeroth order term. Of course suitable gauge-fixing and ghost terms are to be added
to the above action as usual. The quantum effective action or the generator
of the 1-particle-irreducible graphs is then calculated as
\begin{eqnarray}
\no
\Gamma = (\S_{YM} +m^2 \S_m )&+&(\gamma_1 -\Delta^{(1)}) \S_m \\
\no
                              &+&~ (\gamma_2 -\Delta^{(2)})
\S_m +\cdots\\ 
\end{eqnarray}
where the ellipsis stands for higher loop terms, terms with field dependences 
other than the
structure given by $\S_m$ and ghost-dependent and gauge-fixing type terms.
$\gamma_1$ is the coefficient of $\S_m$ generated by the one-loop diagrams with
the rules given by $\S_{YM} +m^2 \S_m$, similarly $\gamma_2$ corresponds to the two-loop
contribution and so on. The $\gamma_i$ are functions of $m$. The parameter $m$ is the
entire dynamically generated magnetic mass by definition, and hence, 
we need the corrections to
vanish. Thus we need $\gamma_1 -\Delta^{(1)}=0,~\gamma_2-\Delta^{(2)}=0$, etc. 
These conditions
determine $\Delta$. Finally, we do not want to change the theory, only rearrange
and resum various terms. Thus $\Delta $ must itself be $m^2$. We thus get
\be
\gamma_1 +\gamma_2 +\cdots =m^2
\ee
This is a `gap equation' which determines $m$ to the order to which the calculation 
is performed.

We have to choose a specific mass term to implement this procedure. 
As mentioned in the introduction, the $\omega_n=0$ mode is the important
one and hence, as a first step, we can use a three-dimensional mass term
and do a three-dimensional calculation. The mass term which 
Alexanian and I have used is given by
\be
S_m=\int d\Omega ~K(A_n,A_{\bar n})
\ee
where $A_n={1\over 2}A_in_i, A_{\bar n}={1\over 2} A_i{\bar n}_i$,
$A_i$ being the gauge potentials.
$n_i$ is a (complex) three-dimensional null vector of the form
$n_i=(-\cos\theta \cos\vf-i\sin\vf,-\cos\theta
\sin\vf+i\cos\vf,\sin\theta)$.
$d\Omega=\sin\theta d\theta d\vf$ and denotes integration over
the angles $n_i$. $K(A_n,A_{\bar n})$ is related to the Wess-Zumino-Witten action
as well as the eikonal for a Chern-Simons theory and is mathematically
very similar to the hard thermal loop effective action. I have discussed elsewhere
the full expression
for
$K(A_n,A_{\bar n})$ and some of its nice geometrical properties \cite{NAN}.
Carrying out the integration over $n_i,{\bar n}_i$,
\begin{eqnarray}
S_m&=&~\int {d^3k \over (2\pi )^3} ~\half
A_i^a(-k) A_j^a(k)\biggl(\delta_{ij}-{k_ik_j\over
{\vec k}^2}\biggr)\no\\ 
&&+\int ~A^a_iA^b_jA^c_kf^{abc}
V_{ijk}\\
V_{ijk}(k,q,p) )&=&  ~-{i\over 6}\delta (k+q+p)
\biggl[{1\over k^2q^2-(q\cdot k)^2}\biggr]\no\\
&&\times\biggl[\biggl\{{q\cdot k\over k^2}-
{q\cdot (q+k)\over (q+k)^2}\biggr\}k_ik_jk_k \no\\
&&~~~+{k\cdot(q+k)\over (q+k)^2}(q_iq_jk_k+ q_kq_ik_j+q_jq_kk_i)\no\\
&&-(q\leftrightarrow k)\biggr]
\end{eqnarray}

The self-consistent procedure then gives, to one-loop order, a gap equation with
the solution \cite{NAN}
\be
m\approx 1.2 ~{e^2N\over 2\pi}
\ee
where $e^2=g^2T$, $g$ being the (3+1)-dimensional coupling constant.

The primary advantage of this approach is that it is perturbatively implementable.
However, some criticisms have also been raised about this approach.
First of all, similar calculations have been done with different mass terms.
Buchmuller and Philipsen have considered a mass term generated by coupling
to a Higgs scalar field \cite{BP}. (A similar mass term was also suggested by
Cornwall \cite{Corn1}.) They find a value which is $(3/4)$-th of our value. Jackiw and Pi
have used a mass term of the form $F_{ij}(D^2)^{-1}F_{ij}$ and find a complex
mass (which is somewhat worrisome) \cite{JP1}. More recently, they have used a mass term
quadratic in the gauge potentials; the gap equation part of the
calculation is then equivalent to a ``unitary
gauge" limit of the Buchmuller-Philipsen calculation and gives the same value 
for the gap \cite{JP2}. The question then arises: 
Does it matter whether different mass terms
give different answers ? Could we take the point of view that one workable
mass term proves the generation of the magnetic mass ? After all, the self-consistent
procedure mixes up different loop orders
when solving the gap equation and hence there is no reason why different mass terms
have to agree to the order calculations have been performed. An optimist might say
that different mass terms could agree in a full calculation. Nevertheless, the
fact that one can get different values for $m$ is worrisome.

Another potential difficulty has been pointed out by Cornwall \cite{Corn2}. 
The residue
$Z$ at the pole of the corrected propagator in our approach is about 180. This is 
a big change from $Z=1$ at the tree level and suggests that one may run into
difficulties, perhaps developing a tachyonic mass, at the two-loop level.
And also, the two-loop corrections are not parametrically small in terms of powers
of coupling. (If they are smaller than the one-loop contribution, it has to be
simply due to numerical factors.) This makes it difficult to get a systematic
expansion and solution for the gap equation, a point which has been emphasized
by Jackiw and Pi \cite{JP2}.
On the positive side though, there is a recent two-loop analysis which
seems to show that the two-loop contribution is only about $15-20\%$ of the
one-loop contribution \cite{eber}.

So where do we stand on this problem at this time ? I believe there are
two distinct questions here. Some of the criticisms do apply to the
estimation of the value of the magnetic mass. To be more confident,
one needs to have some independent nonperturbative analysis, a question to
which I shall return in the next section. However, there is also
the question of including magnetic screening in thermal perturbation
theory. To make the expansion well defined one needs to introduce
an infrared cutoff in a systematic gauge-invariant way. The specific numerical
value is not important to this question of principle; nor is it important
for the practical calculation of some types of processes. It is not consistent
in the course of a perturbative calculation to introduce an ad hoc
infrared cutoff whose value is taken from a lattice or some other nonperturbative
analysis. The self-consistent approach can be useful in such a context.

Finally, there is the question of how much the higher Matsubara frequencies
as well as how much electrostastic screening can modify the result. This has been
analyzed in references \cite{hungary}; 
see also the article by P. Petreczky in these proceedings.

There can be one other way of introducing an infrared cutoff within a perturbative
expansion. This would involve changing the distribution function for the
gauge bosons, a possibility which needs to be explored to see if it can
provide a better systematic scheme. I shall return to this in the last section.

\section{Nonperturbative analysis of $YM_{2+1}$}
\label{sec3}


As I have mentioned in the introduction, the magnetic mass may be considered as the
mass gap of $YM_3$ or $YM_{2+1}$ in a Wick rotated version. Thus analysis of 
$YM_{2+1}$ gives another approach to the magnetic mass. Of course,
the existence of a mass gap in $YM_{2+1}$ or more generally, any kind of
nonperturbative analysis of this theory, is also interesting in its own right.
Here I shall report on a nonperturbative analysis of
$YM_{2+1}$ which we have carried out over the last three years \cite{13,14,15}.
We do a Hamiltonian analysis, starting from first principles; quite simply, what we try
to do is to
write the Schr\"odinger equation for $YM_{2+1}$ and solve it. This is notoriously
difficult in field theory in general, but in our case, we can make significant
progress by using some results from two-dimensional conformal field theory.

Let us consider 
an $SU(N)$-gauge theory in the $A_0 =0$ gauge. The gauge potential 
can be written as $A_i = -i t^a A_i ^a$, $i=1,2$, where $t^a$ are hermitian  
$N \times N$-matrices which form a basis of the Lie algebra of $SU(N)$ with 
$[t^a, t^b ] = i f^{abc} t^c,~~{\rm {Tr}} (t^at^b) = {1 \over 2} \delta ^{ab}$. 
Wavefunctions for physical states must be gauge-invariant and thus they
are functions on the space of all gauge potentials modulo gauge transformations,
i.e., on the gauge-invariant configuration space ${\cal C}$.
It is therefore necessary to make a change of variables to gauge-invariant
quantities. We are not just interested in a formal change of variables. We
will need to calculate explicitly the Hamiltonian in terms of the new variables,
the Jacobian for the change of variables, the volume measure for the inner product of
wavefunctions, etc., and then construct eigenstates of the Hamiltonian.
Thus we need a parametrization of the gauge fields such that these steps can be
explicitly carried out. 
We start by   
combining the spatial coordinates $x_1 ,x_2$ into the complex combinations 
$z=x_1 -ix_2,~{\bar z} =x_1+ix_2$ with the corresponding components for the
potential 
$A\equiv A_{z} = {1 \over 2} (A_1 +i A_2), ~~  
{\bar A}\equiv A_{\bar{z}} = 
{1 \over 2} (A_1 -i A_2) = - (A_z)^{\dagger}$. 
The parametrization we use is then given by 
\be 
A_z = -\partial_{z} M M^{-1},~~~~~~~~~~~~~ A_{\bar{z}} = M^{\dagger -1} \partial_ 
{\bar{z}} M^{\dagger}   
\ee 
Here $M,~M^\dagger$ are complex $SL(N,{\bf {C}})$-matrices. Such a 
parametrization is possible
and is standard in many discussions of two-dimensional gauge fields.  
A particular advantage of this parametrization is the way gauge transformations are
realized. A gauge transformation $A_i \rightarrow 
A_i^{(g)} = g^{-1} A_i g + g^{-1} \partial_i g, ~g(x)\in SU(N)$ is obtained 
by the transformation $M\rightarrow M^{(g)}=g M$. The gauge-invariant degrees of freedom
are parametrized by the hermitian matrix $H=M^\dagger M$, which may be thought of as
elements of $SL(N,{\bf {C}})/SU(N)$. Physical state wavefunctions are functions of
$H$.

I shall now go through the main results of our analysis; details of the calculations
may be found in references \cite{13,14,15}. 
Some of the technical details are also explained in
the article by D. Karabali in these proceedings.
\vskip .1in
\noindent ${\underline {Result ~1}:}$ The volume measure on the gauge-invariant configuration
space (which is the measure of integration for the inner product of states)
can be exactly calculated. It
is given by \cite{13,GKBN}
\be 
\label{measure}
d\mu ({\cal C}) = d\mu (H) e^{2N S(H)}
\ee
\vskip .1in
In this equation, $d\mu (H)$ is the Haar measure for $SL(N,{\bf {C}})/SU(N)$, 
given explicitly
by $d\mu (H) ~= (\det r) [\delta \vf ]$ with $H$ parametrized in terms of a real field
$\vf^a(x)$ and $H^{-1}\delta H~= \delta \vf^a ~r_{ak}(\vf ) t_k $. 
$S(H)$ is the Wess-Zumino-Witten (WZW) action for 
$H$ given by 
\be 
S(H) &&= {1 \over {2 \pi}} \int \Tr (\partial H \bar{\partial} H^{-1})\no\\ 
&&+{i \over {12 \pi}} \int \epsilon ^{\mu \nu \alpha} \Tr ( H^{-1} \partial_{\mu} 
H H^{-1} \partial _{\nu}H H^{-1} \partial _{\alpha}H)  
\ee
This calculation of the measure of integration for the inner products does not suffer
from Gribov or gauge-fixing ambiguities; for more details, see
\cite{13,14,15} and also the article by Karabali in these proceedings.
(Incidentally,
$K(A_n,A_{\bar n})$ in the mass term used in the last section is the same as $S(H)$ 
above with
the identification
$z= x\cdot n,~{\bar z}=x\cdot {\bar n}$ and with all fields being three-dimensional.)

The inner product for states $|1\ra$ and
$|2\ra$, represented by the wavefunctions $\Psi_1$ and $\Psi_2$, is given by
\be
\label{inprod}
\la 1 | 2\ra = \int d\mu (H) e^{2N S(H)}~~\Psi_1^* \Psi_2 
\ee
Observables are ultimately matrix elements of operators, i.e., they are inner products.
By (\ref{inprod}),
this gives a correlator of the two-dimensional WZW model for $H$, which is
a conformal field theory. In other words, we have the exact mapping
$$
\left\{ \matrix{{\rm Matrix~elements~of}&{}\cr
{YM_{2+1}}&{}\cr}\!\right\}= \left\{ \matrix{{\rm Correlators~of~a}&{}\cr
{\rm hermitian ~WZW ~model}&{}\cr}\!\right\}
$$
\vskip .1in
\noindent ${\underline {Result ~2}:}$ Normalizable wavefunctions are functions of the current
$J= (N/\pi ) \partial_zH~H^{-1}$.
\vskip .1in
In other words, the wavefunctions are more restricted than being just functions of
$H$; they can only depend on $H$ via the specific combination in $J$. This result
follows from considerations of integrable representations in the WZW theory via a mapping
between unitary and hermitian models \cite{GKBN}. 
The quantity $J$ may be considered as the 
gauge-invariant definition of a gluon. Since wavefunctions depend only on $J$ by this
result, we can write the Hamiltonian in terms of $J$ and functional
derivatives with respect to
$J$.

This result is also consistent with the fact that the Wilson loop operator,
which provides a complete description of gauge-invariant observables,
can be constructed from $J$ alone as
\be
W(C)= \Tr P ~e^{-\oint_C (Adz+\bA d{\bar z})}=
 \Tr P ~e^{(\pi /c_A)\oint_C J }
\ee
\vskip .1in
\noindent ${\underline {Result ~3}:}$ The $YM_{2+1}$ Hamiltonian is given by
\begin{mathletters}
\be
\label{hamil}
{\cal H}&=& T+V \\
T&=& {e^2\over 2}\int E^a_iE^a_i \no \\
&=& m \Bigl[ \int_u J^a(\vu) {\d \over \d J^a(\vu)}\no \\
&&~~~~~+~ \int \Omega_{ab} (\vu,\vv) 
{\d \over \d J^a(\vu) }{\d \over \d J^b(\vv) }\Bigr] \\
V&=&{1\over 2e^2} \int B^aB^a \no \\
&=&{ \pi \over {m N}} \int \bdel J_a (\vx)
\bdel J_a (\vx) 
\end{eqnarray}
\end {mathletters}
where $m= e^2N/2\pi$ and
\be
\Omega_{ab}(\vu,\vv)&=& {N\over \pi^2} {\d_{ab} \over (u-v)^2} ~-~ 
i {f_{abc} J^c (\vv)\over {\pi (u-v)}}
\ee
\vskip .1in
\noindent ${\underline {Result~ 4}:}$ The kinetic energy operator $T$ obviously has the
ground state wavefunction $\Phi_0 =1$. With the inner product (\ref{inprod}), this is
normalizable. The term $J^a(\vu) {\d \over \d J^a(\vu)}$ in $T$ counts the number of
currents $J$ and each $J$ contributes $m= e^2N/2\pi$ to the eigenvalue.
In particular
\be
T~J^a = {e^2N\over 2\pi} ~J^a 
\ee
This is again an exact result, indicating that we should expect a gap of $m=e^2N/2\pi$ 
for the gluons.
\vskip .1in
\noindent ${\underline {Result ~5}:}$ The vacuum wavefunction $\Psi_0$ is given by
\begin{eqnarray}
\label{wavfn}
\no
\Psi_0= \exp&&\left[ -{1 \over {2 e^2}} \int_{x,y} B_a(\vx) \left[{ 1 \over 
{\bigl( m + 
\sqrt{m^2 - \nabla ^2 } \bigr)} }\right]_{\vx,\vy} \! B_a(\vy) \right]\\
{}&&~~~~~~~~~~+ 3J- ~and ~higher ~J-terms 
\end{eqnarray}
\vskip .1in
$\Phi_0=1$ is the ground state wavefunction for $T$, but it is not an eigenstate
for ${\cal H}=T+V$. By treating $V$ perturbatively, we can obtain a $1/e^2$- or
$1/m$-expansion of the full vacuum wavefunction. Summing up all terms with two powers of
$J$ in $\log \Psi_0$, we get the result 5 \cite{15}.

The first term in (\ref{wavfn}) has the correct (perturbative) high momentum
limit. Thus although we started with the high $m$ or low momentum
limit, the result (\ref{wavfn}) does match onto the perturbative limit.
The higher terms are also small for the low momentum limit.
\vskip .1in
\noindent ${\underline {Result~ 6}:}$ The expectation value of the Wilson loop
is given, for the fundamental representation,
by
\be
\la W_F (C) \ra = {\rm constant}~~\exp \left[ - \sigma {\cal
A}_C \right]
\ee
where ${\cal A}_C$ is the area of the loop $C$ and $\sigma$, which may be identified as the
string tension, is obtained as
\be
\label{tension}
{\sqrt{\sigma }}= e^2 \sqrt{{N^2-1\over 8\pi}}
\ee
\vskip .1in
This is a prediction of our analysis starting from first principles with no adjustable
parameters. Notice that the dependence on $e^2$ and $N$ is
in agreement 
with large-$N$ expectations, with $\sigma$ depending only on the combination
$e^2N$ as $N\rightarrow \infty$.
Formula (\ref{tension}) gives the values $\sqrt{\sigma}/e^2
=0.345, 0.564, 0.772, 0.977$ for $N=2,3,4,5$. 
There are estimates for $\sigma$ based on Monte Carlo simulations of lattice gauge
theory. The most recent results for the gauge  groups $SU(2),~SU(3),
~SU(4)$ and $SU(5)$ are 
$\sqrt{\sigma}/e^2 =$ 0.335, 0.553, 0.758, 0.966 \cite{teper}. We see that
our result agrees with the lattice result to within $\sim 3\%$.

Given this result, we might venture to say that we have a viable approach to
nonperturbative phenomena in $YM_{2+1}$ or $YM_3$.

A comment on the accuracy of our calculation is in order. Why is the result so good ?
The string tension is determined by large area loops and for these, it is the long 
distance part of the wavefunction which contributes significantly.
In this limit, the $3J$- and higher terms in (\ref{wavfn}) are small compared to the
quadratic term. The latter leads to the result (\ref{tension}).
\vskip .1in
\noindent ${\underline {Result ~7}:}$ To lowest order in a power series expansion for the
current
$J\simeq (N/\pi ) \partial \vf_a t_a$, the Hamiltonian can be simplified as \cite{14}
\be
\label{hamil2}
\H \simeq \half \int_x [- {\d ^2 \over {\d \phi _a ^2 (\vx)}} + \phi_a (\vx)  \bigl( m^2 -
\nabla ^2 \bigr)  \phi_a (\vx)] + ...
\ee
where $\phi_a (\vk) \!=\! \sqrt {{N k \bar{k} }/ (2 \pi m)} \vf _a (\vk)$, in momentum space.
\vskip .1in
In arriving at this expression we have expanded the currents and also absorbed the
WZW-action part of the measure into the definition of the wavefunctions, i.e.,
the operator (\ref{hamil2}) acts on 
\begin{eqnarray}
\no
{\tilde \Psi} &&=e^{NS(H)}\Psi\\
&&\simeq e^{-{c_A \over 4\pi} \int
\del \vf \bdel
\vf} \Psi
\ee
The above equation shows that the propagating particles in the
perturbative regime, where the power series expansion of the current is
appropriate, have a mass $m=e^2N/2\pi$. 
This value can therefore be identified as the magnetic
mass of the gluons as given by this  nonperturbative analysis.

We thus see that the self-consistent method, which amounts to selective
summation of terms in the perturbative expansion
gives a value $\approx 1.2 (e^2N/2\pi )$ while the above analysis,
which starts from a high $e^2$-approximation and then connects to the perturbative
limit gives a value $e^2N/2\pi $ for the magnetic mass. The two values are quite
close to each other and also agree with other estimates. I expect that the
actual numerical value of the magnetic mass should be very close to
this number.

Finally, I would like to give a short intuitive argument for why there is a
mass gap, which may help to understand the exact calculations reported above. 
The crucial
ingredient is the measure of integration in the inner product
(\ref{inprod}). 
Writing $\Delta E, ~\Delta B$ for the root mean square fluctuations of the  
electric field $E$ 
and the magnetic field 
$B$, we have, from the canonical commutation rules  
$[E_i^a, A_j^b]= -i\delta_{ij}\delta^{ab}$, $\Delta E~\Delta B\sim k$,  
where $k$ is the 
momentum variable. This gives an estimate for the energy 
\be 
{\E}={1\over 2} \left( {e^2 k^2\over\Delta B^2 } +{\Delta B^2 \over e^2} 
 \right) 
\ee 
For low lying states, we minimize ${\E}$ with respect to $\Delta B^2$,  
$\Delta B^2_{min}\sim 
e^2 k$, giving ${\E}\sim k$. This corresponds to the standard photon. 
For the nonabelian theory, this is inadequate since $\la \H \ra$ 
involves 
the factor $e^{2N S(H) }$. In fact, 
\be
\la \H \ra ~= \int d\mu (H) e^{2NS(H) }~ \half (e^2E^2 +B^2/e^2 ) 
\ee 
In terms of $B$, the WZW action goes like $ S (H) \approx 
[ -(N /\pi ) \half \int B (1/k^2 )B +...]$; we thus see that $B$ follows a Gaussian  
distribution of width $\Delta B^2 \approx \pi k^2 /N$, for  
small values of $k$. This Gaussian dominates near small $k$ 
giving $\Delta B^2 \sim k^2 (\pi /N )$.  
In other words, eventhough ${\E}$ is minimized around $\Delta B^2 \sim k$, 
probability is  
concentrated around $\Delta B^2 \sim k^2 (\pi /N)$. For the expectation 
value of the energy, 
we then find 
${\E}\sim e^2N/2\pi  +{\O}(k^2)$. Thus the kinetic term in  
combination with  
the measure factor $e^{2NS(H)}$ could lead to a mass gap of order $e^2N$.  
The argument is not rigorous, but captures the essence of how a mass gap  
arises in our formalism.

\section{Modifying the Bose distribution}
\label{sec4}


I now return to the question of thermal perturbation expansion for Yang-Mills theory.
The nonperturbative analysis I have outlined can give a good understanding
of the origin of the mass gap and even a value for it.
However, for thermal perturbative $YM$ theory, we would like
 a scheme for magnetic screening which is
perturbatively  implementable. The selfconsistent scheme has this virtue, but is rather
difficult from a calculational point of view. There is also the further difficulty
of the two- and higher loop contributions not being parametrically
smaller. In this
section I propose an alternative method. In the previous analyses, 
we have concentrated on
the spectrum, arguing for a mass gap. The thermal distribution of gluons, which is the
other essential ingredient in any thermal calculation, was unchanged.
It is well known that the `bad' infrared behaviour of the Bose distribution
for gluons of low momenta is what causes difficulties, a situation
similar to Bose-Einstein condensation. In such a situation
there will be
correlations in the ground state. Further, the theory of the mode with
zero Matsubara frequency is equivalent to $YM_3$ at zero temperature with a modified
coupling constant $e^2=g^2T$. Clearly the ground state of $YM_3$
has strong correlations in it.
Thus the distribution of gluons will certainly be modified. 
The proposal is then to start with a
modified distribution  which
incorporates magnetic screening (say depending on a parameter $m$) and then
to determine the parameter by free energy calculations. This is somewhat like a variational
way of determining the distribution function.

The number of ways of distributing $n_k$ particles of
momentum $k$ in
the $G_k$ states of energy $\omega_k =\vert {\vec k}\vert$ is given by
\be
W[\{ n_k \}] = \prod_{k}{(G_k+n_k -1)! \over (G_k-1)! n_k!}
\ee
The maximization of this with equal a priori probabilities for the states and subject to
the condition of fixed
total energy
leads to the standard Bose distribution. In other words, we can determine
the most probable occupation numbers ${\bar n}_k$ by ${\partial /\partial n_k}
\left[ \log W[\{ n_k\}] -\beta (\sum_k n_k\omega_k -E)\right] =0$,
where the term $\beta (\sum_k n_k\omega_k -E)$ enforces the condition of fixed total
energy $E$ via the Lagrange multiplier $\beta$ which is eventually identified as
$1/T$. A simple
way to incorporate magnetic screening is to assign an a priori probability $(1-e^{-\omega
/m})^{n_k}$ to the state and maximize
${\tilde W}= \prod_k (1-e^{-\omega /m})^{n_k} W[\{ n_k \}]$. This leads to the distribution
function
\begin{eqnarray}
\label{distr}
\no
{\bar n}_k &&= {1\over{ e^{\omega /T}-1 + \alpha (\omega )} }\\
\alpha (\omega )&&={e^{\omega /T}\over {e^{\omega /m}-1}}
\end{eqnarray}
${\bar n}_k$ has good infrared behaviour with
${\bar n}_k \rightarrow 0$ as $\omega \rightarrow 0$; further, 
for $\omega \gg m$, ${\bar n}_k$ is exponentially 
close to the standard Bose distribution. The above distribution can be derived from
the density matrix
\be
\label{density}
\rho = e^{-\omega a^\dagger a /T}~e^{-\omega b^\dagger b /m} ~e^{b^\dagger a}
~e^{-a^\dagger b}
\ee
$a^\dagger ,a, b^\dagger ,b$ are the usual creation-annihilation operators.
$a^\dagger, a$ refer to the gluons, $b^\dagger, b$ are similar, but
have no direct meaning; they
are just a mathematical device to obtain (\ref{distr}) in a simple way.

At first glance, there would seem to be many ways of modifying the
a priori probabilities to get magnetic screening. But actually, the choices
are rather limited. The modification (\ref{distr}) has the properties:\hfil\break
1) It has infrared cutoff.\hfil\break
2) Modification is exponentially small in the ultraviolet, so that moment
calculations are not affected.\hfil
\break
3) It is perturbatively implementable.\hfil\break
4) It corresponds to a simple density matrix, so that it is calculationally simple.
\hfil\break
5) It has a gauge-invariant generalization to an interacting theory.

Regarding the last point, a generalization of (\ref{density}) is given by
\be
\rho = e^{-H_{YM}/T} ~e^{-{\tilde H}_{YM}/m}~e^{B^\dagger A}~e^{-A^\dagger B}
\ee
where $H_{YM}$ is the Yang-Mills Hamiltonian, ${\tilde H}_{YM}$ is similar
but constructed out of the $b^\dagger ,b$ operators and $A,B$ can be defined
as eigenopertors of the corresponding Hamiltonians
\be
[H_{YM},A] = -\omega  A,~~~~~~~~~~~[{\tilde H}_{YM},B]= -\omega B
\ee
$A,B$ are a generalization of $a,b$ to the interacting case. By doing free energy
calculations, I expect to determine the parameter $m$ variationally. These calculations
are currently under way.

\acknowledgments

I thank Professor U. Heinz and the other organisers of the 5-th International
Workshop on Thermal Field
Theories and Their Applications-1998 for a nicely organized and useful conference. This
work was supported in part by a grant from the National Science Foundation.


\end{document}